\documentclass[prl,aps,twocolumn,showpacs]{revtex4}
\usepackage{amssymb}
\usepackage{amsmath}
\usepackage{mathptmx}
\usepackage{color}
\usepackage{euscript}
\usepackage{eufrak}
\usepackage{hyperref}
\usepackage{graphicx}
\usepackage{bm}

\def\q{\quad}

\def\equ{Eq.~\eqref}

\newcommand{\eps}{\varepsilon}

\begin{document}

\title{Resonantly suppressed transmission and anomalously enhanced light absorption
\\ in ultrathin metal films}

\author{ I.~S.~Spevak$^1$}
\author{ A.~Yu.~Nikitin$^{1,2}$}
\author{ E.~V.~Bezuglyi$^3$}
\author{ A.~A.~Levchenko$^4$}
\author{ A.~V.~Kats$^1$}
\email{ak\_04@rambler.ru} \affiliation{
 $^1$ A.~Ya.~Usikov Institute for Radiophysics and
Electronics NASU, 61085 Kharkov, Ukraine\\
 $^2$ Departamento de F\'{i}sica de la Materia Condensada-ICMA,
Universidad de Zaragoza, E-50009 Zaragoza, Spain \\
$^3$ B.~I.~Verkin Institute for Low Temperature Physics NASU, 61103 Kharkov, Ukraine\\
$^4$ Department of Physics, University of Minnesota, Minneapolis, MN 55455, USA }

\date{\today}

\begin{abstract}
We study light diffraction in the periodically modulated
ultrathin metal films both analytically and numerically. Without
modulation these films are almost transparent. The periodicity
results in the anomalous effects, such as \emph{suppression of the
transmittance} accompanied by a strong enhancement of the
absorptivity and specular reflectivity, due to excitation of the
surface plasmon polaritons. These phenomena are opposite to the
widely-known \emph{enhanced transparency} of
periodically modulated optically thick metal films. Our theoretical
analysis can be a starting point for the experimental
investigation of these intriguing phenomena.

\end{abstract}

\maketitle

Ten years ago Ebbesen et~al.
\cite{Ebbesen98_Nature} reported on a pioneering observation of
the \emph{enhanced light transmission} through subwavelength hole
arrays. Now being classical, this experiment stimulated
large number of investigations focused on the diffraction
by the optically thick metal films which were
periodically modulated by holes, slits, surface corrugations, etc.
The physical origin of the enhanced light transmission is the
interaction between the eigenmodes of an opaque metal film and the
incoming and outgoing waves~\cite{Ebbesen_theory_PRL01}.
More specifically, periodic modulation leads to
the transformation of incoming photons into surface plasmon
polaritons (SPPs) localized at the film interfaces. The
subsequent back transformation of the excited SPPs into
outgoing photons gives rise to the observed anomalous
effects in the light reflection and transmission.

In this Letter, we predict in some sense the opposite phenomenon. We
consider \textit{ultrathin films} with the
thickness $\ell$ smaller or comparable to the skin depth, $\ell
\lesssim \delta$, which is an opposite case as compared to the
widely studied configurations. For such films, SPPs are separated
into long-range (LR) and short-range (SR) modes
\cite{LRSarid_PRB81,Book_Raether,ThinSambles_PRB08} with the
dispersion relations modified due to the modulation. The light
transmission through the homogeneous ultrathin film is rather high.
We show that a periodic modulation of the
dielectric permittivity of thin films results in the
paradoxical optical effect --- highly-transparent films become
nearly opaque and highly-reflective. The region of such anomalous
behavior is blue-shifted with respect to the position of the
local maximum in the transmittance.
Another amazing feature of the predicted phenomenon is an extraordinary enhanced
absorptivity (up to $50\%$). The latter is pronounced only
in a narrow vicinity of the maximum of the amplitude of the
resonantly excited SPP wave. This {\em resonance point} is situated
between the transmittance maximum and minimum (Fano profile), which are
caused by constructive and destructive interference.

To clarify the most essential physics of the effect, we examine the
simplest case of one-dimensional modulation with symmetric
dielectric surrounding. In our analytical approach, we develop a
resonance perturbation theory based on the small parameter
$\ell/\delta \ll 1$. Analogous analytical approach
(with periodic
modulation supposed to be a small parameter) has been successfully used
before in studies of the resonant diffraction
in periodically modulated metal half-spaces and thick films
\cite{KM,KPS94,Kats_Nikitin_Nesterov_07, Kats_Spevak_07}. Additional
simplifications are provided by large negative values of the real
part of the complex dielectric permittivity $\eps = \eps' +
i\eps''$, $|\eps'| \gg 1$, along with smallness of the absorption,
$\eps'' \ll |\eps'|$. These conditions are typical for noble and
some other metals in visible and near infrared spectral regions.
Finally, our analytical results are supported by the
numerical analysis.

\begin{figure}
\includegraphics[width=7cm]{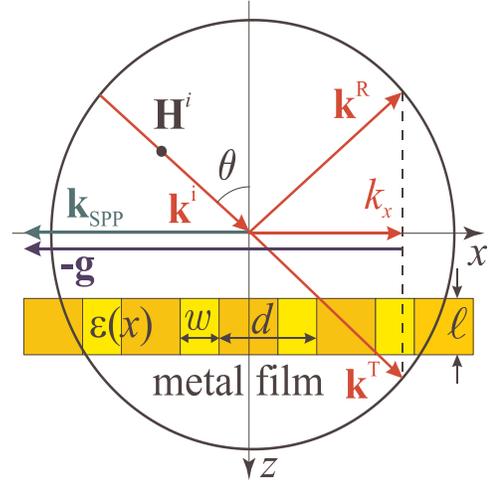}
\caption{(Color online) Geometry of the resonant diffraction
problem. $\mathbf{k}^i$, $\mathbf{k}^R$, and $\mathbf{k}^T$ are the
wave vectors of the incident, specularly reflected, and zero-order
transmitted waves, respectively; $\mathbf{k}_{\mathrm{SPP}}$ is the
wave vector of the excited SPP, $\mathbf{g}$ is the reciprocal
grating vector, $d$ is the grating period, $w$ is the slit
width.}\label{Fig_1}
\end{figure}

Consider a plane monochromatic TM polarized wave, with magnetic
field $\mathbf{H}^i(x,z) = \hat{\mathbf{y}} \exp(i \mathbf{k}^i
\mathbf{r})$, incident onto the periodically modulated metal film
with the grating period $d$, as shown in Fig.~\ref{Fig_1}. The
wavevector of the incident wave $\mathbf{k}^i =
k\sqrt{\eps_d}(\sin\theta, 0,\cos\theta)$, where $k = \omega/c$,
$\eps_d$ is the dielectric permittivity of the surrounding medium,
and $\theta$ denotes the angle of incidence. We represent the
diffracted magnetic field in the half-spaces $z \le 0$, $z \ge \ell$
in the form of the Fourier-Floquet expansion,
\begin{equation}\label{23_04_06_1A}
 H(x,z) = \sum_{n} \exp(iq_{n} x) \left\{
\begin{array}{ll}
  R_n  e^{ i f(q_{n})z} , & z \le 0 \\
  T_n  e^{ -i f(q_{n})(z -\ell)} , & z \ge \ell
\end{array}\right.\,,
\end{equation}
where $f(q) \equiv \sqrt{\eps_d k^{2} - q^{2} }$ with $\mathrm{Re},
\mathrm{Im} f(q) \ge 0$,  $q_{n} = k\sqrt{\eps_d} \sin \theta + ng$, $n = 0, \pm
1, \ldots$, $g=2\pi/d$ is the period of the reciprocal lattice, and
the time dependence, $\exp(-i \omega t )$, is omitted. $R_n $ and $T_n $
are the reflection and transmission
transformation coefficients (TCs), respectively. Inside the film, $0
\le z \le \ell$, the fields are
\begin{eqnarray}\label{9_02_08}
\{H(x,z),\mathbf{E}(x,z)\}=\sum_{n} \{H_{n}(z),\mathbf{E}_{n}(z)\}
\exp(iq_n x  )\,.
\end{eqnarray}
Excluding the normal component of the electric field by means of the
Maxwell equations, we get the set of the first-order linear
ordinary differential equations for the functions $H_n(z)$,
$E_{xn}(z) $. The solution of this set couples the
values of the internal fields on the interfaces. By matching the
fields at the interfaces, we obtain an infinite set of linear
algebraic equations for $R_n $ and $T_n $.

\begin{figure}
\includegraphics[width=8cm,height=12cm]{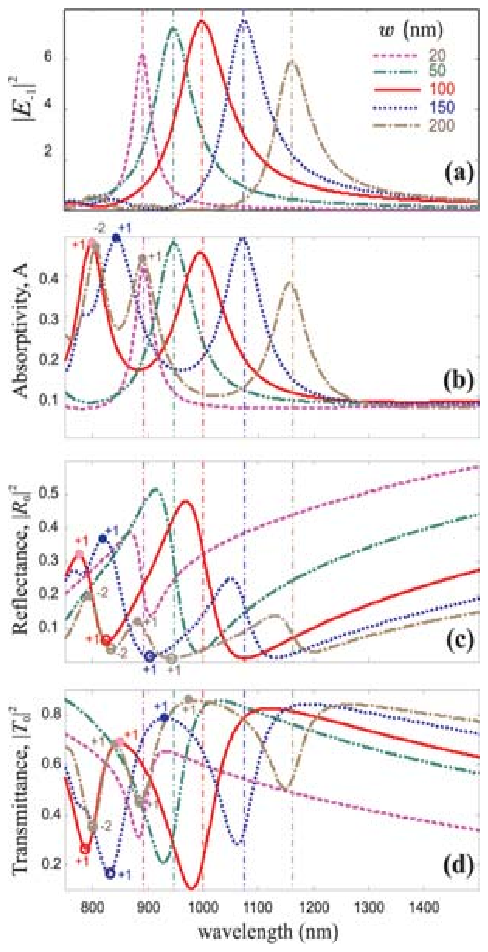}
\caption{(Color online) Wavelength dependence of the $-1$st order
intensity at the film boundary (a), total absorptivity (b), zero
diffraction order reflectance (c), and transmittance (d) (numerical
calculation). The parameters are: $\eps_d =1$,
$\eps_{\mathrm{lamel}}=\eps_{\mathrm{gold}}$, $\eps_{\mathrm{slit}}
= 0.2 \eps_{\mathrm{gold}}$ (the wavelength-dependent gold
permittivity is taken from Ref.~\cite{Moreno-08}), the incident
angle $\theta =45^0$, the grating period $d=400$~nm, the film
thickness $\ell = 10$~nm, the slit width $w$ is shown in the legend
in (a). Vertical dash-dot lines indicate the position of the SR
resonance in $r=-1$ diffraction order; open (filled) circles
indicate minima (maxima) of $|R_0|$ and $|T_0|$ for the $r=1$ and
$r=-2$ resonances. The longest-wavelength extremes (not marked)
correspond to the $r=-1$ resonance.}\label{Fig_2}
\end{figure}

Before proceeding with the analytical treatment, we present an
example of numerical calculations of the resonant spectral
properties of a periodically modulated ultrathin film,
Fig.~\ref{Fig_2}. The calculations have been performed by the
rigorous coupled wave analysis (RCWA) with the improvement suggested
in \cite{Li_96,Lalanne}. We assume an opaque filling of the slits
which restricts the resonances to the SPP modes only and prevents
other resonances (e.g., slit modes). Note that the opaque material
of the filling does not qualitatively influence the results.
Therefore, it is sufficient to chose a high-contrast but opaque
filling resulting in strong resonances even for small slit widths.

Figure~\ref{Fig_2} shows that the transmittance $|T_0|^2$,
absorptivity $A$, and reflectance $|R_0|^2$ possess strong resonant
behavior for various slit widths $w$ in the spectral region typical
for the experiments. Depending on the slit width, the transmittance
in the resonance region is suppressed up to 10\% for $\ell = 10$~nm,
whereas it is of order of 85\% away from the resonance. In contrast,
the reflectance and absorption are greatly enhanced up to 50\%. It
turns out that all shown peculiarities are related to the resonant
excitation of SR SPP in a certain diffraction order. Namely, the
longest-wavelength extremes of $|R_0|^2$, $|T_0|^2$ and $A$
correspond to the $-1$st SR SPP resonance shown in
Fig.~\ref{Fig_2}a, while the short-wavelength extremes are due to
the $+1$st and $-2$nd resonances.

In order to explain quantitatively the resonant
peculiarities of the spectra, we consider
analytically the solution in the vicinity of a single SR resonance.
To this end, we solve the set of equations for the field amplitudes
in the resonant approximation
\cite{KM,KPS94,Kats_Spevak_07,Kats_Nikitin_Nesterov_07} using the
small parameter $\ell/\delta = k\ell\sqrt{|\eps|} \ll 1$. The
resonances correspond to the excitation of both SR and LR
SPP modes. However, as it follows from the analytical considerations and is confirmed by
the computations, just the excitation of
SR SPP strongly affects energy fluxes of the outgoing waves. Within
the second order in $k\ell$, we obtain the resonant TC in
the vicinity of a single SR resonance in $r$th diffraction order,
\begin{equation}\label{11_02_08}
R_{r} =   - \frac{u_{0}}{u_{0} + 2 i} \frac{\mathbb{E}_{r0}}{\mathbb{E}_{00}}
\frac{1  }{F_{r}/F_{SR} - 1 } , \q T_{r} \simeq - R_{r},
\end{equation}
where
\begin{align}
\label{27_10_07_6} F_{m} &= f(q_m)/(k\sqrt{\eps_d}),  \;\; u_{m} = k \ell
\mathbb{E}_{mm} F_{m} , \; \; F_{SR} = 2/ik\ell Z_r,
\\
\label{27_01_08_5}  Z_r &=  \mathbb{E}_{rr} + i k \ell X_{r} , \q
X_{r} =  \sum_{N \ne r} \frac{F_{N} \mathbb{E}_{rN} \mathbb{E}_{Nr}
}{2-iu_N} .
\end{align}
Here $\widehat{\mathbb{E}} = \hat{\zeta}^{-1}$, where the elements
of the T\"oplitz matrix $\hat{\zeta}$ are formed by the Fourier
coefficients of $1/\eps(x)$ expansion, $\zeta_{mn} =
(1/\eps)_{m-n}$, Refs.~\cite{Li_96,Lalanne}.

\begin{figure}[h!]
\includegraphics[width=8cm]{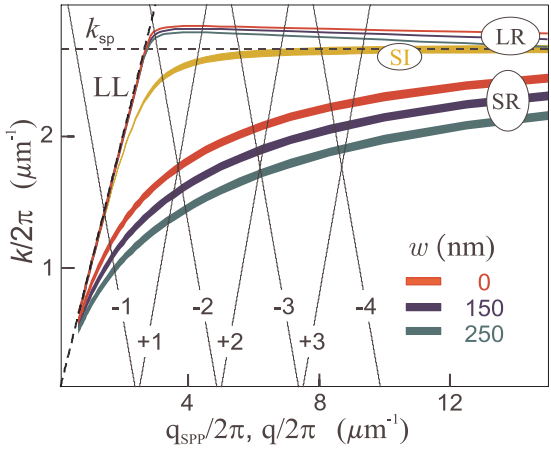}
\caption{(Color online) Dispersion curves for SPPs at the air/gold single
interface (SI), and for the SR and LR SPPs for a homogeneous film with
mean permittivity $\eps_0$ for different slit widths and $\ell
=10$~nm. The thickness of these curves reflects the SPP linewidth.
The light line ($LL$) is shown in black dash. The dashed horizontal
line indicates the surface plasmon frequency, $k_{sp} =
\omega_{sp}/c$, $\eps'(\omega_{sp}) = -1$. Numbered inclined lines
are $q=|q_n| = |k\sin\theta +n g|$ with $n= \pm 1, \ldots$ for
$\theta = 45^\circ$. Intersection of $n$th line with the dispersion
curve corresponds to the resonance excitation of the SPP in $n$th
diffraction order.}\label{Fig_4}
\end{figure}

The resonant TC is proportional to the modulation harmonic
which couples the incident wave with SPP through a one-step scattering.
Other scattering processes contribute to the denominator of
$R_r$. The pole of $R_r$ corresponds to the SR SPP dispersion
relation $q_{SR}^2 \simeq \eps_d k^2 [1 + 4\eps_d (k\ell Z_r)^2]$ modified by
a periodic modulation. This modification is governed by the real part
of $Z_r$, $Z'_r \simeq \eps'_0 - k \ell X''_{r}$, which is
basically determined by the first term. This explains naturally
the red-shift of the resonant peculiarities with
increasing slit width $w$ (see Fig.~\ref{Fig_2}). Indeed,
when $w$ increases, $|\eps'_0|$ decreases and forces the red-shift
of the SPP dispersion relation, as demonstrated in
Fig.~\ref{Fig_4}.

The dissipation and SPP linewidth are governed by the
imaginary part of $Z_r$, $Z''_r \simeq \eps''_0 + k \ell X'_{r}$.
Here the second term exceeds the first one even for a moderate
modulation. Emphasize that the second term is mainly
determined by the summation over outgoing waves [in
\equ{27_01_08_5}] responsible for SPP radiative losses (leakage).

\begin{figure}
\includegraphics[width=8.5cm]{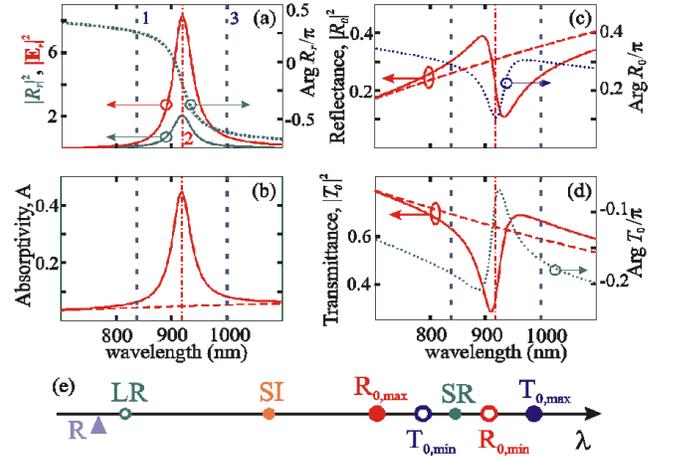}
\caption{(Color online) Results of analytical calculations for
$r=-1$ resonance: wavelength dependence of the resonant
wave magnitude and phase (a), total absorptivity (b), zero-order
reflectance (c), and transmittance (d) for the slit width
$w=100$~nm; other parameters are the same as in Fig.~\ref{Fig_2}.
Vertical lines 1-3 indicate the wavelengths of: SR SPP for slitless
gold film (1) and the SR SPP resonance for modulated film calculated
analytically (2) and numerically (3), cf.~Fig.~\ref{Fig_2}. Dash
lines in (b)--(d) show quantities for a homogeneous film with the
mean permittivity value. The typical ordering of the Rayleigh point,
R, the long-range (LR), single-interface (SI), short-range (SR) SPP
resonances, and the extreme points for the zero-order reflectance
and transmittance are shown in (e). }\label{Fig_1A}
\end{figure}

We show in Fig.~\ref{Fig_1A}a an example of the resonant field
amplitude calculated from
Eqs.~\eqref{11_02_08}--\eqref{27_01_08_5} for $r=-1$.
Due to the excitation of SR SPP,  $R_r$ approaches maximum
in the vicinity of the SPP pole at the point $F_{r} = F_{SR}$ [see
Figs.~\ref{Fig_2}a and \ref{Fig_1A}a]. In view of
$\eps'' \ll |\eps'|$, the  pole is close to the imaginary axis in the
complex $F_{r}$ plane. The condition $\mathrm{Im}[F_r(\lambda,
\theta)] = \mathrm{Im}[F_{SR}(\lambda, \theta)]$ yields the local
maximum of $|R_r|$ in $(\lambda,\, \theta)$ plane. Along this
\emph{resonance line} we have
\begin{equation}\label{1_08_08}
|R_r|_{\mathrm{max}}^2  \simeq \frac{|u_0|^2}{|u_0|^2 +4}
\left|\frac{\mathbb{E}_{r0}}{\mathbb{E}_{00}}\right|^2
\left(\frac{Z'_r}{Z''_r}\right)^2 .
\end{equation}

To study the far field, the knowledge of the
nonresonant-order TCs is necessary. In general, not only
the zero-order but also other diffracted orders  contribute
to the far-field energy flux. Within our analytical approach, they
are given by
\begin{equation}\label{30_01_08_8}
R_{N} = \overline{R}_{N} + \delta R_{N} ,
\q T_N = \overline{T}_N - \delta R_N , \q N \ne r ,
\end{equation}
where $\overline{R}_{N}$, $\overline{T}_{N}$ are contributions from the
nonresonant scattering processes,
\begin{equation}\label{11_02_08_1}
\overline{R}_N = \frac{k\ell F_0 \mathbb{E}_{N0}}{u_N + 2i} ,
\q \overline{T}_N = \delta_{N,0} - \overline{R}_N ,
\end{equation}
and the term $\delta R_{N}$ is the resonant contribution
coming from the SR SPP back-scattering into the
nonresonant diffraction orders,
\begin{equation}\label{27_01_08_8}
\delta R_{N} =  -  k\ell \frac{(F_{r}-ik\ell/2)\mathbb{E}_{Nr}  }{u_N + 2i}
R_{r} .
\end{equation}

For zero-order TCs, $T_0$ and $R_0$, the nonresonant terms coincide
with corresponding Fresnel coefficients $\overline{T}_0$ and
$\overline{R}_0$ for the film with mean value of the dielectric
permittivity, $\mathbb{E}_{00} \simeq \eps_0$. Spectral dependences
of $|\overline{T}_0|^2$ and $|\overline{R}_0|^2$ are shown by dash
lines in Fig.~\ref{Fig_1A}c, \ref{Fig_1A}d. Thus, $T_0$ and $R_0$
result from the interference between the resonant  and direct
(nonresonant) channels. This interference leads to neighboring
minima and maxima in both $|R_0|$ and $|T_0|$ (Fano profile), see
panels (c) and (d) in Figs.~\ref{Fig_1A}, \ref{Fig_2}. According to
Eqs.~\eqref{30_01_08_8}--\eqref{27_01_08_8}, the maximum of
$|T_0|^2$ lies to the right from the above indicated resonant point,
whereas the minimum lies to the left. The maximum of $|T_0|^2$ is
accompanied by the complementary minimum of $|R_0|^2$ and vice
versa. Positions of these complementary extrema are slightly
different, as shown in Fig.~\ref{Fig_1A}e. The resonance also
results in the absorption maximum (in the close vicinity of
$|R_r|^2$ maximum), cf.~Figs.~\ref{Fig_1A}, \ref{Fig_2}. Noteworthy,
even for small active losses, $\eps'' \ll |\eps'|$, the absorption,
being small for ultrathin unmodulated films, increases substantially
(up to 50\%) for an appropriate modulation.

Consider now in detail the case of $r=-1$ resonance. It is
well separated from other resonances at oblique incidence. Moreover,
only zero-order diffracted waves are outgoing and contribute to
the far field. This leads to reduction of the radiative
losses and resonant linewidth as compared with other
resonances. One can see from \equ{1_08_08} and Fig.~2a that  the spectral
maximum of the resonance-order
intensity  depends nonmonotonously on the modulation
(e.g., on the slit width). The intensity is maximal at
$|\mathbb{E}_{r0}|= \mathcal{E}$, when the dissipation losses are equal to
the radiative ones, i.e.,
\begin{equation}\label{1_08_08_1}
\mathcal{E}^2 = \frac{|u_0|^2 +
4}{2 |u_0|} \mathbb{E}''_{00}|\mathbb{E}_{00}| , \q r =-1.
\end{equation}
Substitution of  $\mathcal{E}$ into \equ{1_08_08} yields
$|R_{r}|_{\mathrm{max}}^{2}\rightarrow |R_{r}|_{\mathrm{extreme}}^{2}  \simeq |u_0
\mathbb{E}_{00}|/(8\mathbb{E}''_{00}) \gg 1$. For the modulation level given by
\equ{1_08_08_1}, we obtain $T_0 = R_0 = 1/2$, and the absorptivity
approaches the maximal value $0.5$ in the
resonance. For slit structure with $\eps_d =1$,
$\eps_{\mathrm{lamel}}=\eps_{\mathrm{gold}}$,
$\eps_{\mathrm{slit}} = 0.2 \eps_{\mathrm{gold}}$, and $\theta = 45^{\circ}$,
\equ{1_08_08_1} gives the optimal
value $w/d \simeq 1/4$ which well agrees with the numerical
simulations, cf. Fig.~\ref{Fig_2}.

The minimum value of the transmittance also depends non-monotonously
on the slit width. This minimum can be very low  (specifically, for
transparent slits it may fall to few percents). The minimal value of
the zero-order transmittance is determined by the sum of the
dissipative and radiative losses in nonzero-order diffraction
channels. For $r=-1$ resonance, the nonzero-order radiative losses
are absent, and the transmittance falls up to $|T_0|_{\mathrm{min}}
\propto \eps''_0/|\eps_{r}| \ll 1$.

\begin{figure}[h!]
\includegraphics[width=8.5cm]{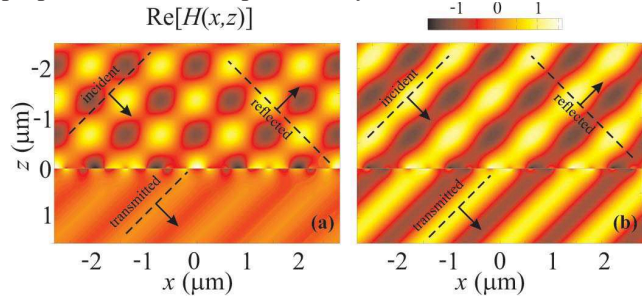}
\caption{(Color online) Spatial distribution of the magnetic field for
$w=100$~nm, $\lambda = 980$~nm in (a), and
$\lambda = 1122$~nm in (b). These wavelengths correspond to the transmittance minimum
and maximum, respectively, for $-1$st order resonance. Other
parameters are the same as in Fig.~\ref{Fig_2}.}\label{Fig_5}
\end{figure}

The predicted effect of the transmittance suppression is illustrated in Fig.~\ref{Fig_5}.
Here we show the instant distribution
of the magnetic field $\mathrm{Re}[H(x,z)]$
for two values of the wavelength related to the minimum and
maximum of the transmittance, cf.~Figs.~\ref{Fig_2}c and \ref{Fig_2}d.
The transmittance
minimum is accompanied by the reflectance maximum. As a result,  the
high-contrast interference pattern appears, see Fig.~\ref{Fig_5}a. On
the other hand, the transmittance maximum is accompanied by the
reflectance minimum resulting in almost complete disappearance of the
interference, Fig.~\ref{Fig_5}b.

In conclusion, we have predicted the paradoxical resonant
properties of ultrathin periodically modulated metal films.
In contrast to widely discussed optically thick films, where the modulation can result
in \emph{extraordinary light transmission},
the periodical modulation of ultrathin films can provide almost \emph{total suppression
of the transmission}. This effect is due to the resonance excitation of the short-range
SPPs, which results also in
up to $50$ percents absorptivity and high reflectivity of the films.

\begin{acknowledgments}
The work was partially supported by STCU grant No 3979.
\end{acknowledgments}

\end{document}